\documentclass[twocolumn,showpacs,preprintnumbers,amsmath,amssymb]{revtex4}
\usepackage{epsfig}

\begin{document}

\title{A new approach to the Sachs-Wolfe effect}

\author{J.T. Mendon\c{c}a$^{1,2}$, R. Bingham$^{2,3}$ and C. H.-T. Wang$^{4,2}$}
\affiliation{$^1$CFP and CFIF, Instituto Superior T\'{e}cnico, 1049-001 Lisboa, Portugal \\ $^2$ CCLRC Rutherford Appleton Laboratory, Chilton, Didcot, Oxon OX11 OQX, U.K.\\ $^3$ Department of Physics, University of Strathclyde, Glasgow G4 0NG, Scotland \\ $^4$ School of Engineering and Physical Sciences, University of Aberdeen, King's College, Aberdeen AB24 3UE, Scotland}


\begin{abstract}
We present a new approach to the Sachs-Wolfe effect, which is based on the dynamics of photons in a space and time varying gravitational field. We consider the influence of plasma dispersion effects on photon propagation, and establish the limits of validity of the usual results of the standard cosmological approach, for the large scale temperature anisotropies of the cosmic microwave background.  New dynamical contributions to the integrated Sachs-Wolfe effect are also discussed.
\end{abstract}


\maketitle

\section{Introduction}

The mapping of the cosmic microwave radiation by satellite observations \cite{smoot,cobe2}, revealed the existence of temperature anisotropies, at various angular scales. It is commonly believed that the large scale anisotropies, of the order of $1$ degree, are due to the Sachs-Wolfe effect \cite{sachs} and that the smaller scale temperature oscillations are associated with the Sunyaev-Zel'dovich effect \cite{suny,peiris}. 

Here we focus on the Sachs-Wolfe effect, which is usually associated with the loss of energy of the photons that, after being freed by the dense baryon plasma in the recombination era, loose energy (or are red-shifted) when they escape from local gravitational wells. These photons can then travel through the universe and carry information about the local confining potentials. This effect is also important in the context of inflation scenarios and for the understanding of dark energy \cite{pogosian}.

Here we propose a new approach to this effect, which allows us to have a clear physical view of the underlying processes and discover additional contributions, not only from the background plasma, but also from dynamical effects associated with the local gravitational perturbations. We notice that, the plasma is a neutral medium, which is associated with perturbations in the density and gravitational potential, as considered in the standard cosmological description, it also contributes to the photon dispersion properties which are usually ignored. This approach is based on the analogy between a gravitational field and a dielectric medium. It is well known that the photon frequency can be up- or down-shifted when light propagates in a space and time varying medium \cite{mendbook}. This effect is called photon acceleration, because the velocity of the frequency shifted photons changes inside the optical medium. This approach was previously explored by us \cite{mendbobpadma} and applied to the study of photon interaction with gravitational waves \cite{mendluke}. We will show here that the Sachs-Wolfe effect is equivalent to photon acceleration and that it results from the photon dynamical properties in non-stationary media.

In Section 2, we establish the equivalent refractive index for a photon in a gravitational field. The influence of a background plasma is considered. The photon equations of motion, written in a canonical form, will then be used as the basic equations for the derivation of the Sachs-Wolfe effect. This effect results from the superposition of several different processes that will be examined separatelly. In Section 3, we study the case of a static gravitational field perturbation in an expanding universe. In this case, we will show that the resulting gravitational red-shift can be seen as pure refraction. The perturbations on the cosmological red shift will also be derived. In Section 4, we describe the dynamical effects associated with a time-varying gravitational perturbation, and obtain new expressions for the integrated Sachs-Wolfe effect. The influence of the plasma, and the contributions of the motion of the gravitational well will be studied. The new additional dynamical contributions to the integrated Sachs-Wolfe are similar to photon acceleration effects already observed in laser experiments \cite{dias,murphy}. They are a direct consequence of the photon dynamics in a space-time varying medium, and cannot  be associated with a simple Doppler effect. Finally, in Section 5, we state our conclusions.

\section{Refractive index of a gravitational field}

In a large variety of situations, the gravitational field changes on space and time scales much larger than the photon wavelength and period. This means that the use of geometric optics is well justified. On the other hand, in general astrophysical conditions, a background plasma component is always present. We can then consider the photon ray equations in a plasma and in a gravitational field. 

The electric field associated with an electromagnetic wave moving in the space and time varying medium (containing both a plasma a gravitational field) can be written as

\begin{equation}
\vec{E} ( x^j) = \vec{A} \exp [ i \psi (x^j) ]
\label{eq:2.1} \end{equation}
where $x^j = (c t, \vec{r}) $ is the four-position, $\vec{A}$  the field amplitude and $\psi (x^j)$ the phase function, or eikonal. The local four-wavevector can then be defined as $k_j = \partial \psi / \partial x^j$, and the local dispersion relation is determined by

\begin{equation}
k_j k^j = g^{ij} k_i k_j = \frac{\omega_p^2}{c^2}
\label{eq:2.2} \end{equation}
where $g^{ij}$ is the metric tensor and $\omega_p$ the local plasma frequency, both depending on the coordinates $x^j$. This dispersion relation implies the existence of a photon effective mass \cite{mendbook}, given by $m_{eff} = \hbar \omega_p / c$. From this equation and from the definition of $k_j$, we can write the Hamilton-Jakobi equation for a photon in a plasma and a gravitational field, in the form

\begin{equation}
g^{ij} \frac{\partial \psi}{\partial x^i} \frac{\partial \psi}{\partial x^j} - \frac{\omega_p^2}{c^2} = 0
\label{eq:2.3} \end{equation}
similar to that of a massive particle in a gravitational field (see reference \cite{landau}). 
Let us now define the photon frequency as 

\begin{equation}
\omega = - c k_0 = - \frac{\partial \psi}{\partial x^0}
\label{eq:2.4} \end{equation}
We can then write the dispersion equation (\ref{eq:2.2}) as

\begin{equation}
\omega = \omega_1 + \sqrt{\nu^2 + \omega_1^2 - \omega_2^2}
\label{eq:2.5} \end{equation}
with
\begin{equation}
\omega_1 = Y^{0 \alpha} k_\alpha c
\quad , \quad \omega_2 = \sqrt{Y^{\alpha \beta} k_\alpha k_\beta} c
\label{eq:2.5b} \end{equation}
where we have used the quantities $\nu^2 =\omega_p^2 / g^{00}$ and $Y^{ij} = g^{ij} / g^{00}$. In these expressions, the greek labels $(\alpha, \beta) =1, 2, 3$ represent the space components. 
Equation (\ref{eq:2.5}) can also be written in a simpler and more appealing form

\begin{equation}
\omega = \frac{k c}{n (k_\alpha, x^\alpha, t)}
\label{eq:2.5c} \end{equation}
where $k = k_\alpha k^\alpha$ is the photon wavenumber, and we have defined the local refractive index of the medium as

\begin{equation}
n (k_\alpha, x_\alpha, t) = \frac{k}{Y^{0 \alpha} k_\alpha} \; \left[ 1 + \nu^2 + \sqrt{1 - \frac{Y^{\alpha \beta} k_\alpha k_\beta}{(Y^{0 \alpha} k_\alpha)^2}}  \right]^{- 1}
\label{eq:2.5d} \end{equation}
This way of writing the photon dispersion relation in a curved space time states the clear analogy between a gravitational field and an optical medium. Even in the absence of a plasma, the curvature of space-time introduces anisotropy and dispersion. 

Equation (\ref{eq:2.5}) shows that the electromagnetic wave frequency $\omega$ is, in general a function of the wavevector $k_\alpha$ and of space and time coordinates $x^j$. This function plays the role of the Hamiltonian in the photon equations of motion. This can be shown by noting that the photon trajectories verify a variational principle $\delta \psi = 0$, which can be explicitly written as

\begin{equation}
\delta \psi = \delta \int k_j d x^j = \delta \int ( k_\alpha d x^\alpha - \omega d t) = 0
\label{eq:2.6} \end{equation}
where we have used $d x^0 = c d t$. This generates the canonical equations

\begin{equation}
\frac{d x^\alpha}{d t} = \frac{\partial \omega}{\partial k_\alpha}
\quad , \quad
\frac{d k_\alpha}{d t} = - \frac{\partial \omega}{\partial x^\alpha}
\label{eq:2.7} \end{equation}
where $\alpha = 1, 2, 3$ and the frequency $\omega = \omega (k_\alpha, x^\alpha, t)$ is the Hamiltonian. In general, this Hamiltonian is time dependent, which implies that the photon frequency will not be a constant

\begin{equation}
\frac{d \omega}{d t} = \frac{\partial \omega}{\partial t}
\label{eq:2.7b} \end{equation}
This means that the photon will loose or gain energy along its trajectory. The same will be true for its proper frequency $\omega'$, which is the frequency actually measured by a local observer. The proper frequency  can be defined as the derivative of the phase function with respect to the proper time $\tau = \sqrt{g_{00}} x^0 c =  \sqrt{g_{00}} t $. This can be stated as \cite{landau} 

\begin{equation}
\omega' = - \frac{\partial \psi}{\partial \tau} = - \frac{\partial \psi}{\partial x^0} \frac{\partial x^0}{\partial \tau} = \frac{\omega}{\sqrt{g_{00}}}
\label{eq:2.8} \end{equation}

The frequency shifts associated with the Sachs-Wolfe effect will be due to particular aspects of photon acceleration and will be induced either by the temporal changes on the metric field $g^{ij}$, or by the temporal dependence of the plasma frequency $\omega_p$. In general, this will imply temporal changes in both $\omega$ and $\omega'$. But that, in some cases, a frequency shift will be measured for a local observer, even when the frequency $\omega$ is constant, as discussed next.

\section{Gravitational and cosmological contributions}

In order to analyze the Sachs-Wolfe effect we study the above photon equations of motion, using the perturbed Robertson metric defined by 

\begin{equation}
d s^2 = (1 + 2 \Phi ) c^2 d t^2 - (1 - 2 \Phi) a^2 (t) d r^2
\label{eq:3.1} \end{equation}
where $\Phi \equiv  \Phi (r, t)$ is the perturbed Newtonian potential, describing a local potential well, and $a (t)$ is the scale factor associated with the expansion  of the universe. We will successively examine three cases, leading to separate contributions to the Sachs-Wolfe effect: 1) the gravitational red-shift directly associated with the local potential well $\Phi (r) \neq 0$, assuming that the scale factor $a (t) = a_0$ is a constant; 2) the cosmological contribution to a local frequency shift, which corresponds to $\Phi = 0$ and local perturbations $\delta a (t)$ of the scale factor $a (t)$; 3) the integrated effect which results from the time dependence of the local Newtonian potential $\Phi (r, t)$. Here we will consider the first two cases, the integrated effects being left to the next Section.

 \bigskip

{\bf i) Gravitational red-shift}

\bigskip

When $\Phi$ is independent of time and the scale factor $a (t)$ is assumed constant,  from the optical point of view we simply have a refractive effect where the photon frequency wave vector changes in time, according to

\begin{equation}
\frac{d k_\alpha}{d t} = - \frac{\partial \omega}{\partial x^\alpha} = \frac{k c}{n^2} \frac{\partial n}{\partial x^\alpha}
\label{eq:3.2} \end{equation}
but where, according to equation (\ref{eq:2.7b}), the photon frequency $\omega$ remains constant. However, the photon frequency $\omega'$ measured by a local observer, as defined by equation (\ref{eq:2.8}), will depend on the observer position. This will lead to an effective frequency shift, as shown next. Assume  that the photon is emitted (or suffers the last scattering) at some position $R$ inside the local gravitational perturbation. Its local frequency would be determined by

\begin{equation}
\omega_e^{'2} = \frac{\omega^2}{g_{00} (R)} = \omega_p^2 (R) + \frac{k_e^2 c^2}{1 - 2 \Phi (R)}
\label{eq:3.3} \end{equation}
where $\omega'_e$ and $k_e$ are the initial values of local frequency and momentum (or wave number). Similarly, if the photon is detected very far away from the local potential, its new local frequency will be given by

\begin{equation}
\omega_f^{'2} = \omega^2 = k_f^2 c^2 = (1 + 2 \Phi (R)) \left[ \omega_p^2 (R) + \frac{k_e^2 c^2}{1 - 2 \Phi (R)} \right]
\label{eq:3.4} \end{equation}
Here we have used the constancy of the Hamiltonian function $\omega$. From these two expressions we can easily obtain

\begin{equation}
\omega'_f \simeq \omega'_e \; \sqrt{1 + 2 \Phi (R)}
\label{eq:3.4b} \end{equation}
which is equivalent to say that, for $| \Phi (R) | \ll 1$, the relative frequency shift measured by a local observer will be $\Delta \omega' \simeq \omega'_e \Phi (R)$. If the photons result from thermal emission, the apparent temperature $T$ of the emitting regions will suffer a variation proportional to this frequency shift, and we get the well know result

\begin{equation}
\frac{\Delta T}{T} =  \Phi (R)
\label{eq:3.4c} \end{equation}
This is the first contribution to the Sachs-Wolfe effect, as perceived by an observer located far away from the emitting region of space-time. Two important aspects should be mentioned. First, this term is independent of the local plasma density, and only results from the changes in the local time scale as determined by $g_{00}$. Second, in terms of geometric optics of space-time, this effect is associated with pure refraction, as shown by the constancy of $\omega$. 

\bigskip

{\bf ii) Perturbed cosmological red-shift}

\bigskip

We now turn to the contributions due to the scale factor $a (t)$. The main argument here is that a local gravitational perturbation $\Phi$ is associated with some local density variation $\Delta \rho$, which also implies a variation of the scale factor $\Delta a$. In order to relate these three quantities, we can start from the continuity equation

\begin{equation}
\dot{\rho} + 3 H ( \rho + p) = 0
\label{eq:3.5} \end{equation}
where $p$ is the pressure and $H = \dot{a} / a$ the Hubble constant. We also know that the density $\rho$ and the pressure are related by the equation of state $p = w \rho$. In a matter dominated universe, we have $w = 0$, and for radiation dominated conditions we get $w = 1/3$. From the continuity equation and from the equation of state we can then write that

\begin{equation}
\frac{\Delta \rho}{\rho} = - 3 (1 + w) \; \frac{\Delta a}{a}  
\label{eq:3.6} \end{equation}

Furthermore, the temperature varies with the scale factor in such a way that $ a T = const$. This well known result could also be derived from the above 
photon ray equations. This means that $a \Delta T = - T \Delta a$. From where we conclude that

\begin{equation}
\frac{\Delta T}{T} = \frac{1}{3 (1 + w)} \; \frac{\Delta \rho}{\rho} 
\label{eq:3.6b} \end{equation}

It can also be realized that, by definition,  the Newtonian potential $\Phi$ is proportional to $- \Delta \rho$, and we can identify $2 \Phi = - \Delta \rho / \rho$. This finally leads to an additional contribution to the Sachs-Wolfe effect, as given by

\begin{equation}
\frac{\Delta T}{T} =  - \frac{\Delta a}{a}  =  - \frac{2 \Phi(R)}{3 (1 + w)}
\label{eq:3.7} \end{equation}
This result was previously derived in reference \cite{white} by following a similar but distinct procedure. Therefore, the two effects determined by equations (\ref{eq:3.4c}) and (\ref{eq:3.7}) will partially cancel each other, and we get for the  temperature fluctuations the resulting value of 

\begin{equation}
\frac{\Delta T}{T} =    \frac{1 + 3 w}{3 (1 + w)} \; \Phi(R)
\label{eq:3.7b} \end{equation}
For a matter dominated universe ($w = 0$) this would lead to the well known expression of $\Delta T / T =	¤ \Phi (R) / 3$.
 
\section{Dynamical contributions}
 
 Let us now consider the case where the scale factor $a(t)$ is assumed constant, but both the local potential well and the plasma frequency depend on space and time, $\Phi (r, t)$ and $\omega_p (r, t)$. The integrated Sachs-Wolfe effect will appear as a particular case of the the total dynamical contributions to the apparent temperature fluctuations, as shown next. The photon equations of motion, assuming one dimensional propagation, are now given by
 
\begin{equation}
\frac{d r}{d t} = c \left[ \frac{1 + 2 \Phi}{1 - 2 \Phi} \right]^{1/2} \; \frac{1}{\sqrt{1 + \omega_p^2 (1 - 2 \Phi) / k^2 c^2}}
\label{eq:3.8} \end{equation}
and

\begin{equation}
\frac{d k}{d t} = - \frac{\partial \omega}{\partial r} = -  \frac{2 k^2 c^2}{\omega (1 - 2 \Phi)^2} \frac{\partial \Phi}{\partial r} - \frac{(1 + 2 \Phi)}{2 \omega} \frac{\partial \omega_p^2}{\partial r} 
\label{eq:3.8b} \end{equation}
We see that there are two kinds of forces acting on the photon and changing its momentum. They result from the gradients of the potential $\Phi$ and of the plasma density. Assuming that both $\omega_p^2$ and $\Phi$ have a similar variation length, we can say that only for low plasma densities (or high local temperatures) such that
$\omega_p^2 \ll 4 k^2 c^2  | \Phi |$ will plasma effects be negligible. In this limit of a negligible plasma density, and for small potential perturbations $| \Phi | \ll 1$, equation (\ref{eq:3.8b}) can be written in a much simpler form

\begin{equation}
\frac{d k}{d t} = - \frac{\partial \omega}{\partial r} \simeq - 2 k c  \frac{\partial \Phi}{\partial r}
\label{eq:3.9} \end{equation}
which has the solution

\begin{equation}
k (t) = k (0) \exp \left[ - 2 c \int_{r(0)}^{r(t)} \frac{\partial \Phi (r(t'), t')}{\partial r} d t' \right]
\label{eq:3.10} \end{equation}
In the limit of small frequency shifts, this can be further approximated to

\begin{equation}
k (t) = k (0) \left( 1 - 2 \int \frac{\partial \Phi}{\partial r} ds \right)
\label{eq:3.10b} \end{equation}
where we have used $d s = c dt$. In general, the total change in wavenumber results from both refraction effects and photon acceleration. But, when plasma effects are neglected, it can easily be seen that $\delta \omega / \omega = \delta k / k$ and that refraction in the present context cannot occur. This means that, in terms of the observation on Earth of thermal photons emitted from the perturbed regions of space time, this results in apparent temperature variations $\Delta T / T = \Delta \omega' / \omega'(0)$, that can be approximated (to the lowest order in $\Phi$) by $\Delta k / k(0)$, where $\Delta k = k (t) - k(0)$, as

\begin{equation}
\frac{\Delta T}{T(0)} = - 2 \int_{r(0)}^{r(t)} \frac{\partial \Phi}{\partial r} ds
\label{eq:3.11} \end{equation}
This expression is the well known integrated Sachs-Wolfe effect, and was obtained here as a limiting case of equation (\ref{eq:3.10}), which was itself an approximated version of the photon equations of motion. It should be noticed here that, if we retain the exact equations, and in contrast with the case of a stationary gravitational perturbation $\Phi = \Phi (r)$, we have in general additional contributions to this effect and due to the plasma medium. These new contributions are not included in equation (\ref{eq:3.11}), but are contained in the exact photon equations (\ref{eq:3.8}) and (\ref{eq:3.8b}). Such plasma contributions can be easily understood, in the following way.  Inside the plasma, photons move with a lower velocity, thus taking a longer time to travel across the local perturbation $\Phi$. They will therefore perceive a stronger potential variation. In addition to slowing down photons, the gradients of the plasma density will create a new force acting on the photons and changing their energy. Therefore, we will expect two different plasma contributions, which will be discussed next.

The dynamical contributions to the integrated Sachs-Wolfe effect can be particularly significant when the gravitational perturbation moves with a finite velocity $u$, and is not simply related with the amplitude variation of a fixed spacial profile. In order to illustrate these new aspects we assume that the perturbed potential takes the form $\Phi (r, t) = \Phi (r - u t)$. It can easily be shown \cite{mendbook} that the quantity 
$\Omega = \omega (\eta) - u k$, with $\eta = r - u t$, is a dynamical invariant. Therefore, for a photon trajectory corresponding the the following initial conditions $\Omega = \omega (0) - u k_e$, we can write for its final properties at detection $\Omega = \omega (\eta_f) - u k_f$. On the other hand, the two values of the photon frequency are determined by

\begin{equation}
\omega (0) = k_e c  \left( \frac{1 + 2 \Phi_0}{1 - 2 \Phi_0} \right)^{1/2} \sqrt{1 + \omega_{p0}^2 (1 - 2 \Phi_0) / k_e^2 c^2}
\label{eq:3.12} \end{equation} 
where $\Phi_0 = \Phi (0)$ and $\omega_{p0}$ are the local potential perturbation and plasma frequency at the emitting region, and $\omega (\eta_f) = k_f c$. Using the invariance of $\Omega$, and assuming a one-dimensional photon propagation along the coordinate $r$, we obtain for the initial and final proper frequencies $\omega'_e = \omega' (0)$ and $\omega'_f = \omega' (\eta_f)$, as defined by equation (\ref{eq:2.8}), the following relation  
 
\begin{equation}
\omega'_f = \frac{\omega'_e}{(1 - \beta)}  \left[ \sqrt{1 + 2 \Phi_0} - \beta \sqrt{1 - 2 \Phi_0} \left(1 - \frac{\omega_{p0}^2}{\omega^{'2}_e} \right)^{1/2} \right]
\label{eq:3.13} \end{equation}
where $\beta = u / c$.  This expression can be simplified by assuming a small potential perturbation $| \Phi| \ll 1$, and a low density plasma such that $(\omega_{p0} / \omega'_e)^2 \ll 1$. In this case we obtain an expression for the corresponding relative temperature fluctuation $\Delta T / T = \Delta \omega' / \omega'$, given by 

\begin{equation}
\frac{\Delta T}{T} \simeq \frac{ \Phi_0}{(1 - \beta)} \left[ 1 + \beta \left( 1 - \frac{\omega_ {p0}^2}{2 \omega^2_T} \right) \right] + \frac{\beta \omega_{p0}^2}{2 \omega^2_T (1 - \beta)}
\label{eq:3.14} \end{equation}
where we have used the local thermal frequency such that $\hbar \omega_T = k_B T$.

This results generalizes equation (\ref{eq:3.4c}) for the case of a moving cloud, where the plasma effect is included. Notice that, in the static case treated above, the existence of a background plasma frequency only introduced refractive effects and did not contribute to the photon frequency shifts, or to the resulting apparent temperature perturbations. By taking $\beta = 0$ indeed we recover equation (\ref{eq:3.4c}). But the existence of a cloud motion changes everything. Not only a finite $\beta$ changes the value of the temperature perturbations associated with the local potential perturbation $\Phi$, as shown by the first two terms, but it also results into an additional plasma contribution, given by the last term in equation (\ref{eq:3.14}). This term is independent of the gravitational field and only results from the plasma motion. This purely plasma effect has been identified before, in the case of laser interactions with time varying plasmas, and studied in various experimental configurations \cite{dias,murphy}. We notice that even a very small plasma frequency, such that  $\omega_p^2 \sim \omega_T^2 \Phi_0 / \gamma^2 \ll \omega_T^2$, will give a contribution comparable to (\ref{eq:3.4c}),  for moderate values of the relativistic gamma factor of the moving cloud, $\gamma \sim 1 / \sqrt{(1 - \beta)}$. On the other hand, for purely vacuum gravitational perturbations, with negligible plasma, $\omega_{p0} \simeq 0$, equation (\ref{eq:3.14}) reduces to
 
\begin{equation}
\frac{\Delta T}{T} \simeq \Phi_0 \frac{ (1 + \beta)}{(1 - \beta)}
\label{eq:3.14b} \end{equation}
For $\beta = 0$ we get equation (\ref{eq:3.4c}), but for high velocities $\beta \rightarrow 1$ this will tend to $\Delta T / T \simeq 2\Phi /(1 - \beta)$, which shows that even a very small potential perturbation can eventually induce a very high temperature fluctuation. This could be relevant to relativistic jets, which are very common astrophysical objects.
  
It is important to notice that the new contributions to the integrated Sachs-Wolfe effect described by equation (\ref{eq:3.14}) cannot be identified with a simple Doppler effect, because they  depend on both the amplitude of the gravitational potential $\Phi_0$, and the local plasma frequency. Therefore, in contrast with Doppler shifts,  the effects described by equation (\ref{eq:3.14}) are genuine dynamical effects and not  kinematic ones. 

\section{Conclusions}

We have proposed here a new approach of the Sachs-Wolfe effect, which is responsible for the large perturbations of the observed cosmic radiation background. This approach is based on photon dynamics and explores the analogy between a gravitational field and an optical medium. The present dynamical approach is particularly well adapted to include additional contributions to the apparent local temperature oscillations, due to the existence of a plasma background. 
We have derived the simple Sachs-Wolfe, as well as the integrated effect, and obtained the usual expressions as limiting cases. We have also identified new dynamical contributions, resulting from the motion of the potential wells in the emitting (or scattering) region that cannot be identified with simple Doppler effects. This work provides a unifying approach that can be useful for a more detailed and accurate modeling of the local structures of the universe, which are associated with the observed oscillations of the cosmic temperature map.

\bigskip

{\bf Acknowledgements}

This work was partially supported by the Centre for Fundamental Physics at the Rutherford Appleton Laboratory.

\end{document}